\documentclass[a4paper]{elsarticle}
\usepackage{amsmath}
\usepackage{amssymb}
\usepackage{graphicx}
\begin{document}
\begin{frontmatter}
\title{Anti-rumor dynamics and emergence of the timing threshold on complex network}
\author[label1,label2]{Kaihua Ji}
\author[label1]{Jiawei Liu}
\author[label1]{Gang Xiang\corref{cor1}\fnref{fn1}}
\cortext[cor1]{Corresponding author}
\fntext[fn1]{Fax(O): (86)-28-85412323, Tel(O): (86)-28-85416478}
\ead{gxiang@scu.edu.cn}
\address[label1]{College of Physical Science and Technology, Sichuan University, Chengdu 610064, China}
\address[label2]{Wu Yuzhang Honors College, Sichuan University, Chengdu, 610064, China}
\begin{abstract}
Anti-rumor dynamics is proposed on the basis of rumor dynamics and the characteristics of anti-rumor dynamics are explored by both mean-field equations and numerical simulations on complex network. The main metrics we study are the timing effect of combating rumor and the identification of influential nodes, which are what an efficient strategy against rumor may concern about. The results indicate that, there exists robust time dependence of anti-rumor dynamics and the timing threshold emerges as a consequence of launching the anti-rumor at different delay time after the beginning of rumor spreading. The timing threshold as a critical feature is further verified on a series of Barab\'{a}si-Albert scale-free networks (BA networks), where anti-rumor dynamics arises explicitly. The timing threshold is a network-dependent quantity and its value decreases as the average degree of the BA network increases until close to zero. Meanwhile, coreness also constitutes a better topological descriptor to identify hubs. Our results will hopefully be useful for the understanding of spreading behaviors of rumor and anti-rumor and suggest a possible avenue for further study of interplays of multiple pieces of information on complex network.
\end{abstract}
\begin{keyword}
Anti-rumor dynamics \sep The timing threshold \sep Rumor control strategies \sep Complex network
\end{keyword}
\end{frontmatter}
\section{Introduction}
Nowadays, with the fast development of instant messengers and SNS networks, information (such as news, and opinions) has the possibility to spread rapidly to a large-scale and may have strong social influences. It is thus significant to understand the dissemination of rumors and their counterparts, anti-rumors, in social networks. Since an authoritative source (for instance, government, large corporation, etc.) for one person may be untrusted for another, online social networks play unprecedented key roles in broadcasting both rumors and anti-rumors because people tend to trust their friends more. Thus, after abstracting real social networks to online topological networks, we can operate quantitative analysis on information propagation by taking advantage of dynamical perspective [1,2]. In this paper, we will introduce the anti-rumor dynamics, analyze it by using mean-field equations and extensive numerical simulations on complex network, and explore how to design efficient strategies for combating rumors in a network-scale.

Similar to those spreading behaviors that occur in various complex networks [3], such as propagation of epidemic diseases in social networks and cascading failure in power grids, rumor dynamics is proposed to analyze the rumor spreading in communication networks quantitatively [4,5], where only single piece of information is considered. However, since multiple pieces of information exist simultaneously in real-world networks, there are basic problems that have not been answered yet, such as, how do the rumors and anti-rumors interplay with each other? If rumors are hazardous, how can we prevent them from further spreading? Trying to answer these questions, on the basis of rumor dynamics we propose the anti-rumor dynamics, in which the anti-rumor is launched at a certain delay time after the beginning of rumor spreading. Meanwhile, anti-rumor dynamics also takes the interaction between the propagations of rumor and anti-rumor into account. Specifically speaking, anti-rumor's propagation has impact upon rumor's propagation and vice versa through two mechanisms: anti-rumor's priority over rumor after the acceptance of anti-rumor, and the prior hypothesis bias before the acceptance of anti-rumor. We shall come back to these two mechanisms later on. In doing so, we can explore several quantities characterizing the dynamics of anti-rumor spreading process, among which the timing threshold that emerges as a consequence of launching the anti-rumor at different delay time is a critical quantity. It is a characterization factor of the delay time that separates two regimes. Since the timing threshold is a network-dependent feature, we will first show that it can be observed explicitly in a communication network and then investigate this quantity systematically in a series of Barab\'{a}si-Albert scale-free networks (BA networks) with the fixed network size. The communication network [6] used in this paper covers all the email communication within a dataset of around half million emails, which contains $N=36,691$ nodes with the average degree $<k>=10.02$. This is a complex network having the feature of power-law distribution [7,8].

The paper is organized as follows. In Section 2, we elaborate the model of anti-rumor dynamics based on rumor dynamics. Next in Section 3, we use both mean-field equations and numerical simulations to explore the key factors in the design of efficient strategy for combating rumor: the timing effect and the identification of influential nodes. Finally, the results are summarized in Section 4.
\section{ANTI-RUMOR DYNAMICS}
Considering if there only exist the rumor and anti-rumor in a network, the basic characteristics of anti-rumor dynamics are as follows. Firstly, the time when anti-rumor is broadcasted is later than that of rumor. If we record rumor's outset as $t=0$ in a network, then the time of launching an anti-rumor is $t=T_{in}$, with $T_{in}>0$. Secondly, both rumor and anti-rumor are essentially in the form of messages, so they follow the same rules of message spreading accordingly. Thirdly, based both on actual situations and on the belief that nothing is more convincing than the truth, we assume that the anti-rumor possesses the priority over rumor, which means a node that has accepted the anti-rumor as true will not believe in the rumor anymore [9]. We call this mechanism after the acceptance of anti-rumor as ``anti-rumor¡¯s priority over rumor" (APOR). There is another mechanism concerning about the action of rumor on the anti-rumor before the acceptance of anti-rumor will be discussed later in Section 3. The characteristics mentioned above suggest that anti-rumor dynamics should be established on the basis of rumor dynamics. Next, we will first give a brief introduction to the classic rumor dynamics (CR dynamics) [10], and then try to build the mechanism of anti-rumor on the basis of CR dynamics in this section.

In CR dynamics, each of the elements of the network can be in three possible states. We call nodes holding a rumor and willing to transmit it {\sl spreaders}, nodes that are unaware of the rumor called {\sl ignorants}, while those that already know it but are not willing to spread the rumor anymore are called {\sl stiflers}. The three states can be represented with {\sl S, I, R} respectively. The density of {\sl ignorants, spreaders, stiflers} at time $t$ are denoted as $i(t), s(t), r(t)$ separately such that $i(t)+s(t)+r(t)=1$, $\forall t.$ The initial conditions are set such that $i(t)=1-1/ N, s(t)=1/N, r(t)=0$ when  $t=0$. At each time step, {\sl spreaders} contact all the neighboring nodes. When the {\sl spreader} contacts an {\sl ignorant}, the latter one turns into a new {\sl spreader} with a possibility $\alpha$.  On the other hand, the {\sl spreader} becomes a {\sl stifler} with a possibility $\lambda$ if a contact with another {\sl spreader} or a {\sl stifler} takes place. The dynamics terminates when $s(t)=0$.

Recent research has observed the absence of influential spreaders in rumor dynamics, which is contrary to the case of epidemic dynamics [2,11]. However, another analysis of real data [12] indicates that influential nodes do exist in rumor dynamics and the larger the coreness of a node is, the more important it is in the network and the closer it is to the center of the network. Intuitively speaking, the influence of information is relevant with the prestige, class and social influence of its promulgators. In consideration of these results, it is necessary to make some amendments to CR dynamics. The modification as follows is proposed [13] to identify the influential nodes. The possibility for an ignorant to believe in the rumor, i.e., to turn into a {\sl spreader} or a {\sl stifler}, still is $\alpha$. However, not all of the {\sl ignorants} that believe in the rumor will decide to diffuse it and become {\sl spreaders}, and they have a possibility $(1-p)$ to turn into {\sl stiflers}. Thus the modified dynamics can be described as: when a {\sl spreader} contacts an {\sl ignorant}, the latter turns into a {\sl spreader} with probability $\alpha$$p$ and into a {\sl stifler} with probability $(1-p)$$\alpha$. The modified dynamics is called rumor dynamics with parameter $p$(PR dynamics).

The anti-rumor dynamics we consider here is built on the basis of PR model, which is therefore called as APR dynamics for short. Since both rumor and anti-rumor are essentially in the form of messages, we also classify anti-rumor's nodes into {\sl ignorants'}, {\sl spreders'} , and {\sl stiflers'}, which can be called {\sl I', S', R'} respectively, according to their awareness levels of anti-rumor. We denote the density of {\sl I', S'}, and {\sl R'} at time $t$ as $i'(t), s'(t), r'(t)$ respectively, with $i'(t)+s'(t)+r'(t)=1$, $\forall t$. And the initial conditions are set such that $i'(t)=1-1/N, s'(t)=1/N, r'(t)=0, t=T_{in}$.

After one node that acts as the promulgator diffuses the anti-rumor in a network, rumor spreading is no longer running freely. For convenience, either {\sl S'} or {\sl R'} is contained in {\sl F}, whose density is $f(t) \equiv s'(t)+r'(t)$. Because a node cannot accept both rumor and anti-rumor at the same time and of anti-rumor's priority, {\sl I, S} and {\sl R} can only be in the {\sl I'} class while {\sl F} cannot become {\sl I, S} or {\sl R} anymore. In APR model the interactions amidst nodes can be denoted in the following formulas:
\begin{equation}
\left\{ \begin{array}{ll}
I\xrightarrow[S]{p\alpha}S \\
I\xrightarrow[S]{(1-p)\alpha}R \\
S\xrightarrow[S, R \,or \,F]{\lambda}R \\
I'\xrightarrow[S']{p\alpha'}S'    & ,t\geq T_{in}\\
I'\xrightarrow[S']{(1-p)\alpha'}R' \\
S'\xrightarrow[S' \,or \,R']{\lambda'}R' \\
\end{array} \right.
\end{equation}
where the phrases under the arrowheads are the contact conditions while those over the arrowheads are the possibilities of transitions. When $p=1$, these formulas will fade into the formulas in ACR model, which is the model of anti-rumor dynamics set up on the basis of CR model. Additionally, the dynamics in APR model will not terminate until $s(t)=0$ and $s'(t)=0$.

We then carry out numerical simulations for both of APR and ACR models on the email network previously introduced. For each simulation, the seed of a rumor is fixed. Without losing generality we set rumor's seed at a node whose degree is equal to the average degree of network. At $t=T_{in}$ a node acting as an {\sl S'} in the network first spreads the corresponding anti-rumor while the other nodes belong to {\sl I'} class. $A=100$ times of simulations have been repeated for every node, i.e., every vertex of the network acts as the initial {\sl spreader} of anti-rumor $A$ times in order to obtain statistically significant results. In this way, we average the final density of {\sl R} for each node in the network as:
\begin{displaymath}
r_{\infty}^{i}=\frac{1}{A}\sum_{m=1}^A r_{\infty}^{i,m}
\end{displaymath}
where $r_{\infty}^{i,m}$ represents the final density of {\sl R} for a particular run $m$ with the seed of anti-rumor at node i. $r_{\infty}^{i}$ quantifies the impact of promulgator i when combating a rumor. Therefore, we can obtain $M_{k}$, which represents the average density for all runs with a seed of $k$ degree, and $M_{k_{s}}$, which represents the average density for all runs with a seed of $k_{s}$ coreness:
\begin{displaymath}
M_{k}=\sum_{\substack{i\in \gamma_{k}}} \frac{r_{\infty}^{i}}{N_{k}},
M_{k_{s}}=\sum_{\substack{i\in \gamma_{k_{s}}}} \frac{r_{\infty}^{i}}{N_{k_{s}}}
\end{displaymath}
where $\gamma_{k}$ stands for the set of all $N_{k}$ nodes with $k$ values and $\gamma_{k_{s}}$ represents the set of all $N_{k_{s}}$ nodes with $k_{s}$ values. Similarly, the final density of {\sl R} obtained when node i acts as the initial {\sl spreader} is averaged over all the seeds with the same $k$ and $k_{s}$ values. So we get $M_{k,k_{s}}$:
\begin{displaymath}
M_{k,k_{s}}=\sum_{\substack{i\in \gamma_{k,k_{s}}}} \frac{r_{\infty}^{i}}{N_{k,k_{s}}}
\end{displaymath}
where $\gamma_{k,k_{s}}$ is the set of all $N_{k,k_{s}}$ nodes with $k$ and $k_{s}$ values.

\begin{center}
\includegraphics[scale=0.5,bb=0 0 500 280]{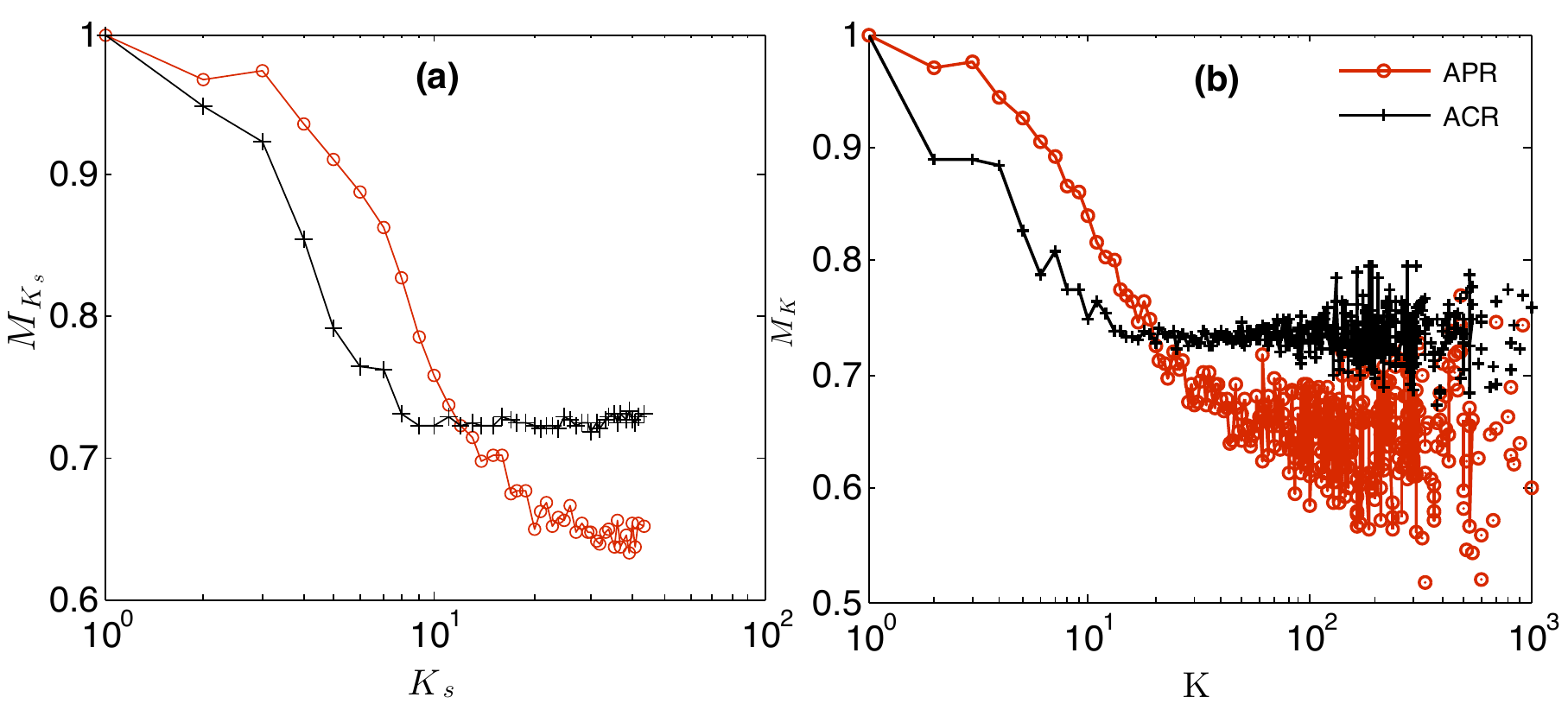}
\end{center}
{\scriptsize
FIG. 1. The comparison between the numerical simulation results from APR and ACR models. We fix rumor's initial spreader and vary the promulgator of anti-rumor. The parameters are $\alpha=\alpha'=0.5,\lambda=0.1$ and $p=0.1$. We get (a)$M_{k_{s}}$ and (b)$M_{k}$ for both ACR and APR models as a function of the class of anti-rumor's promulgator. For visual consideration, $M_{k}$, $M_{k_{s}}$ are normalized.}

From Fig. 1, we find that anti-rumor's impact can be partially evaluated by degree or coreness of its seed in APR model while the absence of influential spreaders takes place in ACR model: in APR model, the larger a node¡¯s coreness is, the more influential it is in the network, which leads to a smaller $r_{\infty}$ at the same time; in ACR model, $M_{k}$ and $M_{k_{s}}$ change lightly after $k_{s}>10$ or $k>10$, which is on the contrary with the practical data [12]. Associating with disease propagation, the alike difference is applied to the cases of epidemical dynamics and CR dynamics as well [11,14]. In fact this phenomenon can be interpreted theoretically by the concept of "spontaneous state transition" [1]. In conclusion, on one hand, PR model introduces a kind of modification which makes itself closer to the actual situation compared with CR model. On the other hand, Fig. 1 shows that APR model is more useful in distinguishing influential spreaders, which means closer to the practical data. Therefore, we will mainly adopt APR model in the following sections.

\section{RESULTS AND DISCUSSIONS}

The delay time of launching the anti-rumor after the beginning of rumor spreading, or the timing effect, is one of the most important aspects that we study for anti-rumor dynamics. In the following series of numerical simulations, we fix the initial spreader of the rumor. Under such a circumstance, the delay time $T_{in}$ and the seed of anti-rumor are the independent variables, while $r_{\infty}^{T_{in}}$ is the dependent variable. So we can get:
\begin{equation}
r_{\infty}^{T_{in}}=\frac{1}{N}\sum_{n=1}^{N}\frac{1}{A}\sum_{m=1}^{A} r_{\infty}^{T_{in},m,n}
\end{equation}
where $r_{\infty}^{T_{in},m,n}$ represents the final density of {\sl R} for a particular run m with a specific anti-rumor's seed n and a specific $T_{in}$. And $r_{\infty}^{T_{in}}$ quantifies how deep an anti-rumor that starts from $T_{in}$ can prevent a rumor from penetrating the network. A more efficient anti-rumor process would correspond to a smaller $r_{\infty}^{T_{in}}$ value.

APR model contains the APOR mechanism which only considers the effect of anti-rumor on rumor after the acceptance of anti-rumor. In analogy to action and reaction forces, this APOR mechanism only accounts for ``reaction force'', and we need to take the corresponding ``action force" into account. Based on actual situations, it is reasonable to assume the nodes that have believed in the rumor prior to the anti-rumor may have smaller possibilities to accept the anti-rumor and become {\sl I'} or {\sl R'} than those {\sl ignorants} of rumor. This mechanism before the acceptance of anti-rumor is called ``the prior hypothesis bias'' (PHB). With considering the PHB mechanism, APR model is changed to APR-PHB model. APR-PHB model can be described by adjusting the fourth and fifth formulas in Eq.(1) to the following form:
\begin{equation}
I'\left\{ \begin{array}{ll}
if \,\, I:\xrightarrow[S']{p\alpha_{1}}S' \\
if \,\, S \,or \,R:\xrightarrow[S']{p\alpha_{2}}S' \\
if \,\, I:\xrightarrow[S']{(1-p)\alpha_{1}}R' \\
if \,\, S \,or \,R:\xrightarrow[S']{(1-p)\alpha_{2}}R'
\end{array} \right.
\end{equation}
where $\alpha_{1}>\alpha_{2}$. Without losing generality, we set the parameters as $\alpha=0.4$, $\lambda=0.2$, $p=0.2$ for APR model, and $\alpha_{1}=0.4$, $\alpha_{2}=0.2$, $\lambda=0.2$, $p=0.2$ for APR-PHB model in the following sections.

\subsection{The timing threshold in communication network}

\begin{center}
\includegraphics[scale=0.5,bb=0 0 400 320]{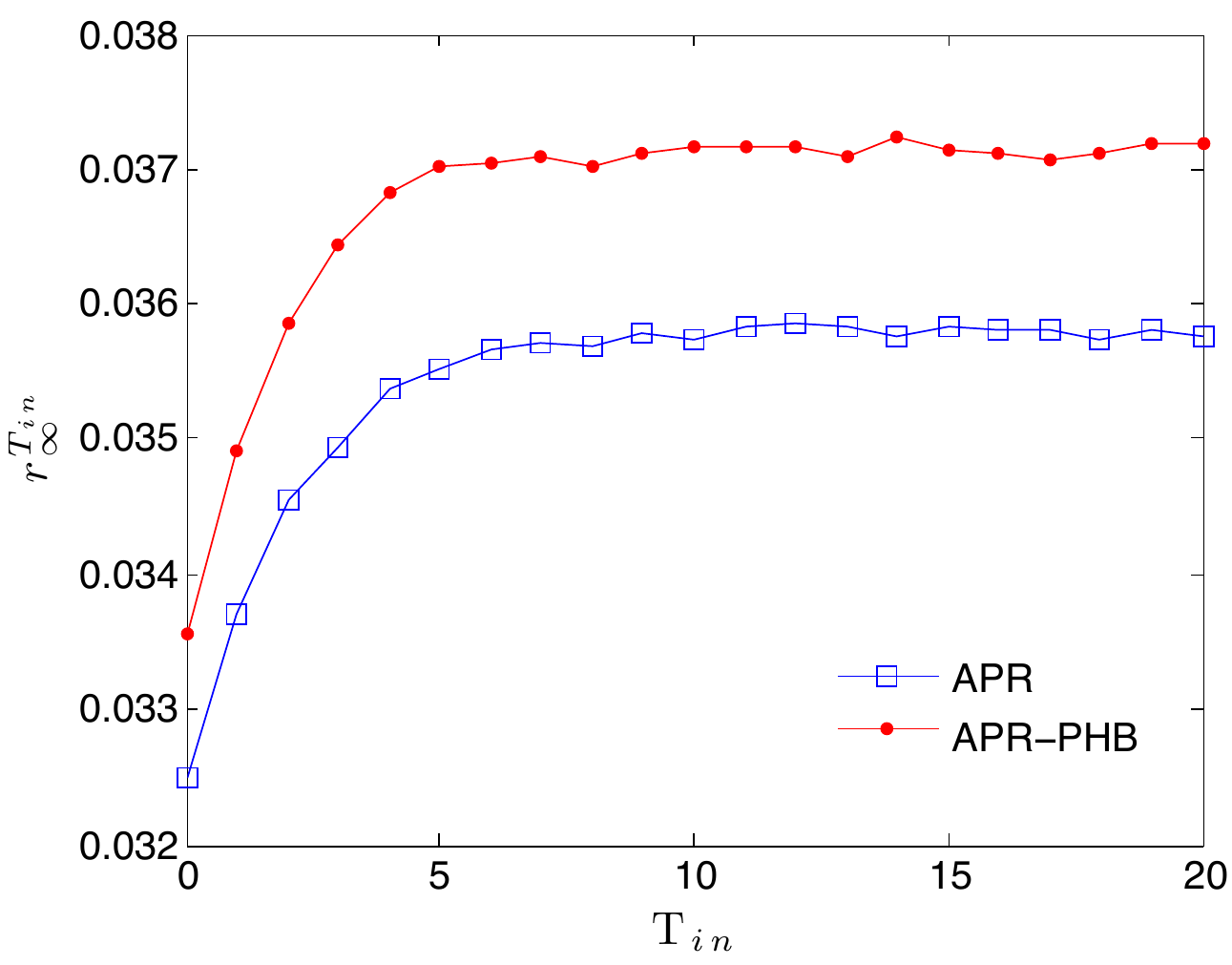}
\end{center}
{\scriptsize
FIG. 2. Growth of {\sl R}'s final density as a function of the delay time $T_{in}$ in an email network. Curves show that the timing threshold emergences for
both APR and APR-PHB models, which is equal to the delay time $T_{0}$ when the asymptotic value of $r_{\infty}^{T_{in}}$ is reached. The results indicate that the timing threshold is approximately equal to $T_{0}=6$ for both APR and APR-PHB models.}

As shown in Fig. 2, the simulation results indicate that the timing threshold emergences in the email communication network for both APR and APR-PHB models. The timing threshold $T_{0}$ is a characterization factor of the delay time $T_{in}$ between two regimes. In the first regime where $T_{in}<T_{0}$, {\sl R}'s final density $r_{\infty}^{T_{in}}$ increases as $T_{in}$ increases, which indicates that we can more efficiently prevent the rumor from spreading by launching the anti-rumor with a shorter delay time $T_{in}$; while in the second regime where $T_{in}>T_{0}$, the evolution of $r_{\infty}^{T_{in}}$ is in equilibrium and the asymptotic values are reached, which indicates that the timing effect is no longer a determining factor for combating rumor.

The fluctuations in Fig. 2 come from the sources as follows. Since the method for studying the timing effect described above is a stochastic method, the result can be randomly unstable and thus influence the observation of the timing threshold.  However, we can control the fluctuation of the result to a certain extent by increasing the simulation times. For Fig.2, $A=10$ times of simulation were carried out. In fact we can obtain relatively stable results by just repeating a moderate number of simulation times with implementing Eq.(2). This is because in each simulation every node in network acts as the promulgators for once and the result has been averaged already.

In addition, there is another aspect worth taking into account. $T_{in}$ is a discrete variable in numerical simulation and the interval between two neighboring discrete points of $T_{in}$ is 1, while the delay time should be a continuous variable in reality. This disadvantage of anti-rumor dynamics causes that we cannot determine the timing threshold more precisely than the order of $O(10^{0})$ in terms of the time in APR and APR-PHB models.

\subsection{The timing threshold in Barab\'{a}si-Albert network}

Since the timing threshold is a network-dependent quantity, we can further scrutinize the timing threshold by looking at its value in other complex networks, for instance the networks generated by algorithms, through the same stochastic method described above in Eq.(2). Most of the real-world networks share some common topological properties [14]. Among those properties, the power-law degree distribution is perhaps the most crucial one [7,8]. In this section we study the timing threshold in the networks generated by BA model that have power-law degree distribution.

\begin{center}
\includegraphics[scale=0.5,bb=0 0 400 280]{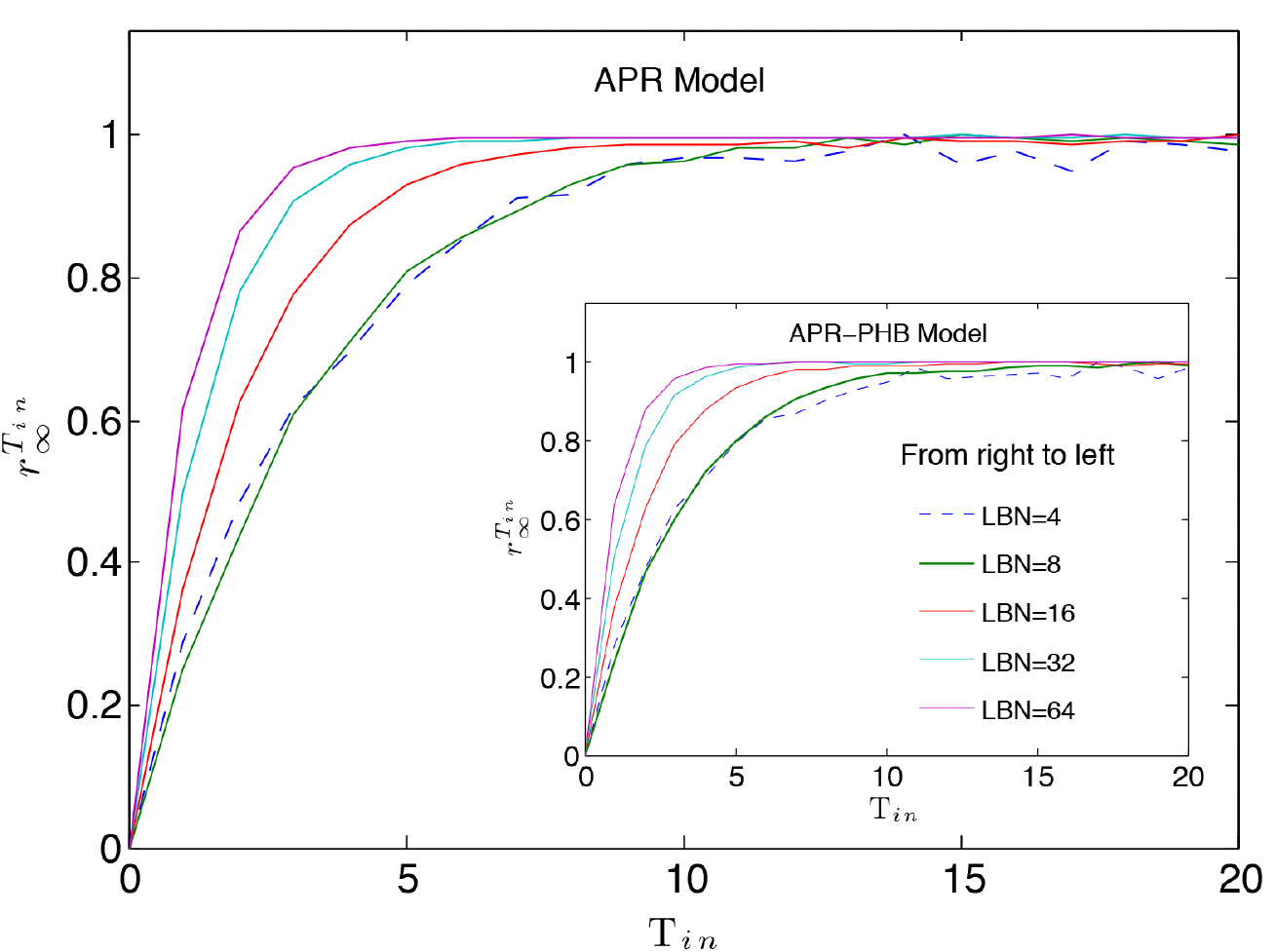}
\end{center}
{\scriptsize
FIG. 3. Normalized $r_{\infty}^{T_{in}}$ as a function of $T_{in}$ for several networks generated by the BA model with a fixed number of nodes $N=3000$ and links by new nodes (LBN)=4, 8, 16, 32 and 64. The timing thresholds can be observed explicitly in all BA networks for both APR and APR-PHB models in this figure, especially for those networks with large LBN. The curves indicate that the timing threshold decreases as LBN increases until close to zero. The times of simulation are $A=100$ for all cases.}

As expected, the timing threshold arises explicitly in BA network as shown in Fig. 3. There are two input parameters when generating a BA network: the number of nodes, and links by new nodes (LBN) [8]. Since the number of nodes is fixed at $N=3000$, LBN is the only variable parameter for generating such a series of BA networks. Therefore, Fig. 3 also shows the characteristics of the timing threshold in a more generalized manner, i.e., the timing threshold is a network-dependent quantity in BA networks, which decreases as LBN increases until close to zero. Note that the average degree of the network is directly proportional to LBN for BA networks with the same size [15]. Thus the negative correlation between the timing threshold and LBN can also lead to the conclusion that the timing threshold tend to come into being more quickly in a denser BA network. Since knowing the timing threshold can help us to decide the timing of launching the anti-rumor, this conclusion is of great practical importance for the design of efficient strategy.

\subsection{Mean-field equations in homogeneous network}

In the meanwhile, the timing effect can also be demonstrated by using mean-field equations derived from formulas (1):
\begin{equation}
\left\{ \begin{array}{ll}
\dot{i'}=-\alpha'ki'(t)s'(t) \\
\dot{s'}=p\alpha'ki'(t)s'(t)-\lambda'ks'(t)[s'(t)+r'(t)] \\
\dot{r'}=(1-p)\alpha'ki'(t)s'(t)+\lambda'ks'(t)[s'(t)+r'(t)] \\
\dot{f}=\alpha'ki'(t)s'(t) \\
\dot{i}=-\alpha ki(t)s(t)-\alpha'ki'(t)s'(t)i(t)/[1-f(t)] \\
\dot{s}=p\alpha ki(t)s(t)-\lambda ks(t)[s(t)+r(t)+f(t)] \\ -\alpha' ki'(t)s'(t)s(t)/[1-f(t)] \\
\dot{r}=(1-p)\alpha ki(t)s(t)+\lambda ks(t)[s(t)+r(t)+f(t)] \\ -\alpha' ki'(t)s'(t)r(t)/[1-f(t)]
\end{array} \right.
\end{equation}
Eq.(4) is derived in the case of homogeneous network [10]. It can be extended to the heterogenetic network in essentially the same form.

\begin{center}
\includegraphics[scale=0.5,bb=0 0 400 300]{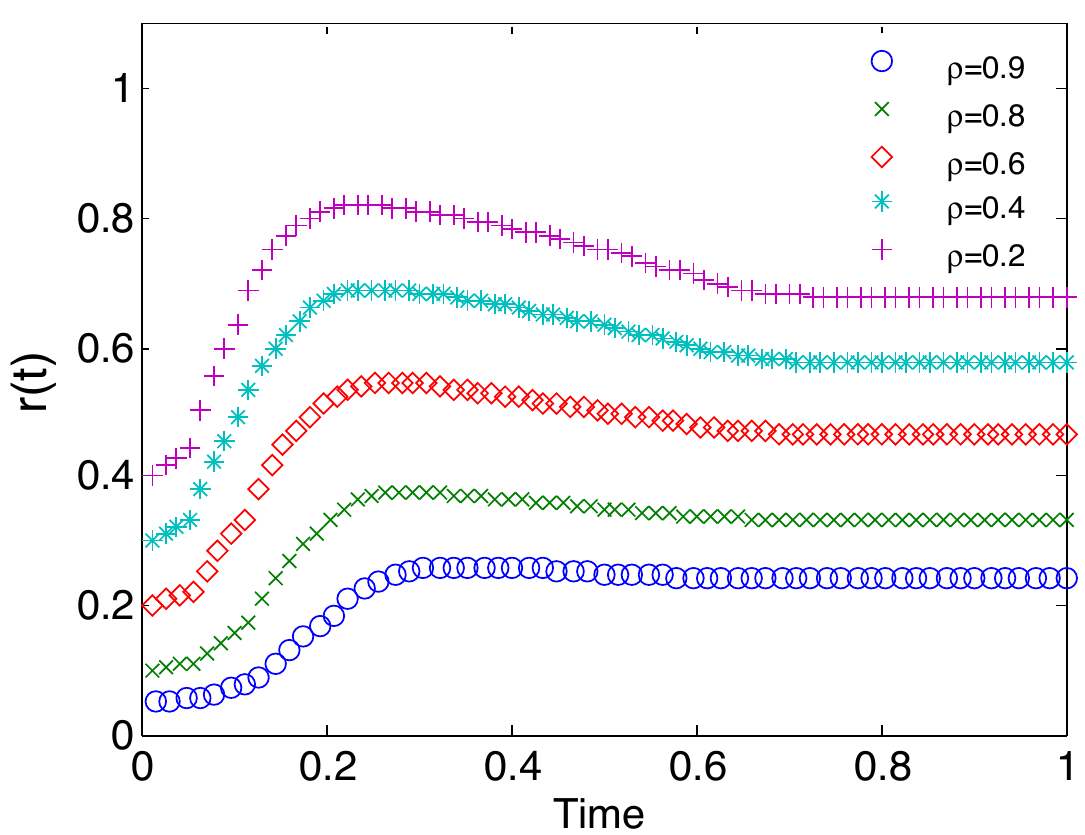}
\end{center}
{\scriptsize
FIG. 4. The computational solutions of the mean-field equations of anti-rumor dynamics. The value of $\rho$ is the density of the {\sl ignorants} of rumor in the network at $t=T_{in}$. Thus a larger $\rho$ corresponds to a smaller $T_{in}$, which represents the situation of broadcasting anti-rumor at an earlier time. The time is normalized. The results show that all of the curves will be stabilized in the end and the decreasing of curves' asymptotic values suggest that the shorter the delay time is, the better effect for combating rumor it will get.}

We compare the situations of launching anti-rumor with different delay time by giving the mean-field equations with different initial conditions. As shown in Fig. 4, each curve represents one situation, in which we will finally obtain an asymptotic value of $r(t)$ in equilibrium for each solution. From the computational results we can get the conclusion that the earlier to broadcast the anti-rumor, the better effects we could get, which is exactly the part of the conclusions that we obtain from previous numerical simulations in the first regime $T_{in}<T_{0}$. However, the timing threshold is not shown in Fig.4, because the mean-field equations are the anti-rumor dynamics derived based on homogeneous network, which reduces many-body problem into one-body problem on one hand, but loses details about the topological properties of complex network on the other hand.

\subsection{The influential promulgators of anti-rumor}

Besides the timing effect, we also focus on how the topological location of a seed can influence anti-rumor's spreading. Since betweenness centrality is proved to be ineffective in distinguishing hubs [2], we will mainly analyze which one, the degree or coreness, is better at predicting a node¡¯s influence in anti-rumor dynamics. This analysis may be helpful in improving the dissemination of civil opinions or restraining the diffusion of hazardous information in complex networks.

\begin{center}
\includegraphics[scale=0.5,bb=0 0 400 320]{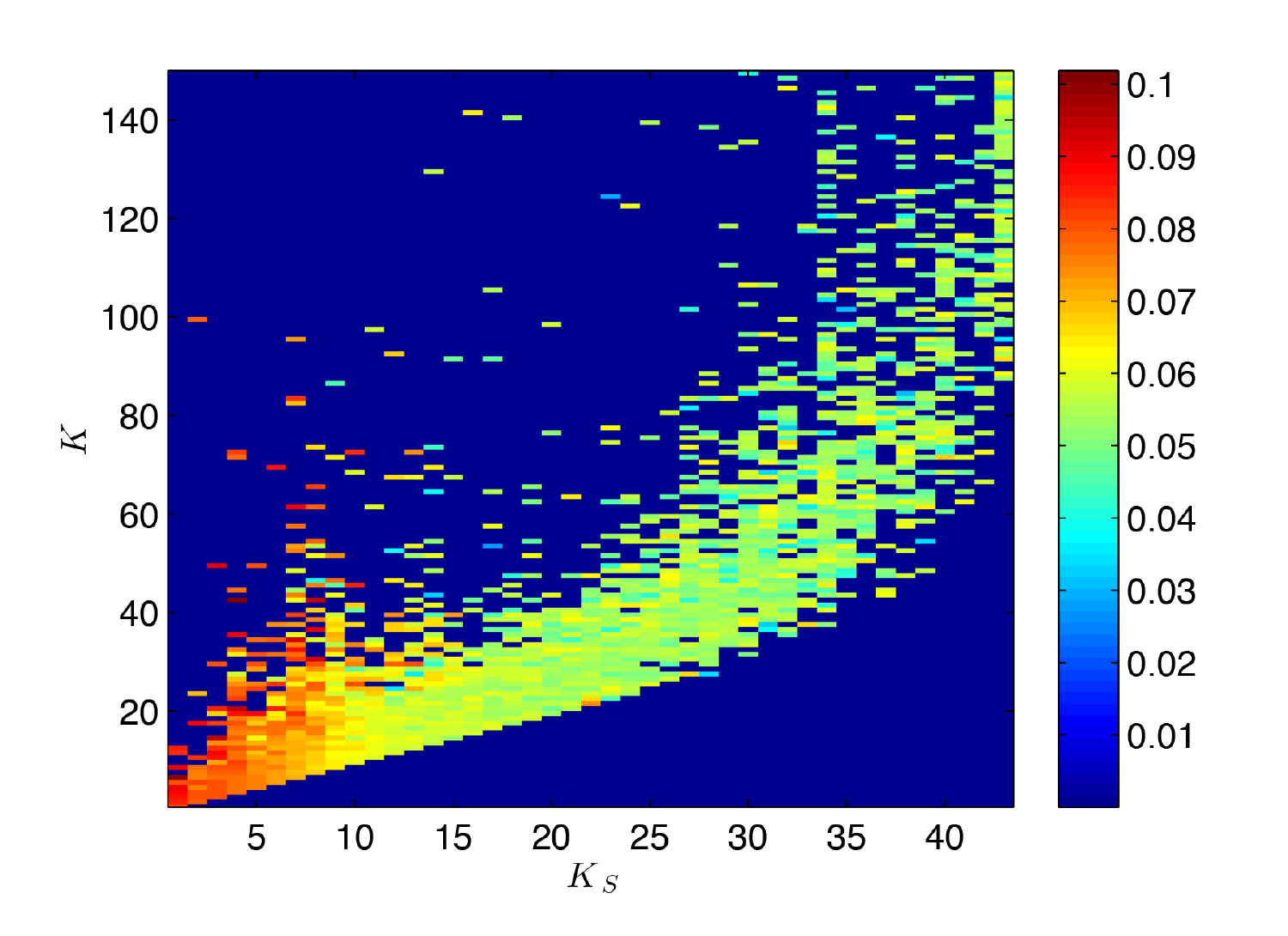}
\end{center}
{\scriptsize
FIG. 5. The comparison between the reliability of degree and coreness for predicting the influence of anti-rumor's promulgator. The color of class $(k, k_{s})$ represents awareness level $M_{k,k_{s}}$, namely, the percentage of nodes that believe in the rumor in the end. For visual purpose, we omit the segment of $k>150$. The results are explained in the text. (For interpretation of the references to color in this figure legend, the reader is referred to the web version of the article.)}

In Fig. 5, we compare the reliability of degree and coreness for denoting a node's influence in anti-rumor dynamics. When $k$ is fixed as some $k_{0}$ the color of the line $k=k_{0}$ changes observably as $k_{s}$ increases. However, when we fix $k_{s}$ as some $k_{s0}$, the color of the line $k_{s}=k_{s0}$ keeps relatively steady as $k$ changes. The larger the value of a node's coreness is, the closer the node is to the top of a hierarchical structure of the network, and the greater influence for combating rumor it holds. Therefore, coreness also constitutes a better topological descriptor to identify influential nodes in anti-rumor dynamics, which is consistent with previous research studying the dynamical process on complex networks [2].

Based on the results, an efficient rumor control strategy should take advantage of the timing effect and the influential promulgator of anti-rumor in the overall planning. Specifically speaking, if the anti-rumor is able to be broadcasted before the timing threshold, then the earlier the better, but if not, then the priorities should be given to other factors like the persuasiveness of information, i.e., the value of $\alpha$ and $p$, but not the timing anymore. What is more, if the case is in BA network, we should keep in mind that the timing threshold tends to come into being earlier in a network with larger average degree. However, since most of the real-world networks have many other topological properties besides the power-law degree distribution, how to precisely predict the timing threshold for a certain real-world complex network still needs further investigations. On the other hand, under the same conditions, the influential nodes identified by large coreness are the better promulgators for anti-rumor, so they are what the strategy should give priorities to.

\section{CONCLUSIONS}

In this paper, we have studied the anti-rumor spreading process on complex network by taking two mechanisms of interactions between rumor and anti-rumor into account: APOR after the acceptance of anti-rumor, and PHB before the acceptance of anti-rumor. By implementing both APR and APR-PHB models to complex networks, we have carried out extensive numerical simulations to study the identification of influential anti-rumor spreaders, and a critical network-dependent quantity---the timing threshold, which characterizes the dynamics of anti-rumor spreading process in a certain network. The results show that the timing threshold $T_{0}$, as a characterization factor between two regimes of anti-rumor's delay time, emerges in a real-world communication network for both APR and APR-PHB models. In the meanwhile, the feature of the timing effect in the first regime has also been verified by the mean-field equations in homogeneous networks. We then extend the method to the case of BA network. As expected, the timing thresholds arise explicitly on a series of BA networks, and a more generalized conclusion has been made accordingly: if the network size is fixed, the timing threshold is a network-dependent quantity in BA networks for both APR and APR-PHB models, which decreases as the average degree of the network increases until close to zero. The characteristics of the timing threshold studied here is of great practical importance since the timing effect plays a key role in the design of efficient strategies against rumor. The same kind of strategies may also consider the identification of influential promulgators and coreness is also shown to be a better indicator.

In both APR and APR-PHB models, the best precision of the timing threshold is of the order $O(10^{0})$ because of the discrete delay time and the fluctuation that comes from the stochastic method. Since it is possible to improve the precision, of further interests would be a more careful exploration of the determination of the timing threshold in various complex networks, so that we can predict the timing threshold in real-world complex networks more precisely. In this sense, our results are useful for the understanding of anti-rumor spreading process and the design of efficient strategies for combating rumor. The methods in this paper may also suggest a new path for further study of interplays of multiple pieces of information on complex network.

\begin{center}
{\bf ACKNOWLEDGEMENTS}\\
\end{center}

This work was supported by Program for New Century Excellent Talents in University (NCET) through Grant No. 11-0351 and by Project Sponsored by the Scientific Research Foundation (SRF) for the Returned Overseas Chinese Scholars (ROCS) of State Education Ministry of China. And we thank Wang Hao for discussions.

\end{document}